\documentclass[a4paper,11pt]{article} \usepackage[a4paper,top=3.5cm,bottom=3.5cm,left=2.999cm,right=2.999cm]{geometry}
\usepackage[english]{}
\usepackage[utf8]{inputenc} 
\usepackage{graphicx}
\usepackage{amsthm,amssymb,bbold, amsmath}
\usepackage{amsfonts}
\usepackage{amsmath}
\usepackage{xcolor}
\usepackage{mathrsfs} 
\usepackage{hyperref}
\hypersetup{hidelinks}

\usepackage{cite}
\hypersetup{
    colorlinks=true,
    linkcolor=blue,
    filecolor=magenta,      
    urlcolor=blue,
     citecolor=magenta,
    pdftitle={Overleaf Example},
    pdfpagemode=FullScreen,
    }
\usepackage{authblk}
\title{Peer influence breaks ergodicity in an opinion dynamics model with external information}
\author[1,2]{Federica De Domenico}
\author[3,4,5]{Fabio Caccioli}
\author[1,2,3]{Giacomo Livan}
\affil[1]{Dipartimento di Fisica, Universit\`a degli Studi di Pavia, Via A. Bassi 6, 27100, Pavia, Italy}
\affil[2]{Istituto Nazionale di Fisica Nucleare, Sezione di Pavia, Via A. Bassi 6, 27100, Pavia, Italy}
\affil[3]{Department of Computer Science, University College London, 66-72 Gower Street, London WC1E 6EA, United Kingdom}
\affil[4]{Systemic Risk Centre, London School of Economics and Political Science, London WC2A 2AE, United Kingdom}
\affil[5]{London Mathematical Laboratory, 8 Margravine Gardens, London WC 8RH, United Kingdom}
\begin{document}
\maketitle

\begin{abstract}
We present a stochastic imitation-based model of opinion dynamics in which agents balance social conformity with responsiveness to an external signal. The model captures how populations evolve between two binary opinion states, driven by peer influence and noisy external information. Through both memory-less and memory-based implementations, we identify a critical threshold of social sensitivity that separates an ergodic phase—where agents collectively track the external signal—from a non-ergodic phase characterized by persistent consensus and reduced adaptability to external changes. Analytical results and simulations reveal that memory in decision-making smooths the transition and lowers the critical threshold for ergodicity breaking. Extending the model to various network structures confirms the robustness of the observed phase transition. We further discuss empirical methodologies for estimating the critical threshold and show how the model may be applied to real-world domains. Our findings contribute to understanding how social conformity, memory effects and randomness jointly shape collective behaviour, with implications for predicting social tipping points and influencing large-scale social dynamics.
\end{abstract}

\section{Introduction}
Opinion formation is rarely an isolated process. Social life is inherently interactive and the beliefs individuals adopt are often shaped by a combination of peer influence, social context, personal convictions and external information. Agreement with others leads to social approval, reinforcing existing opinions, while disagreement can induce social discomfort~\cite{homans1986fifty}. This interplay between internal preferences and social feedbacks forms the foundation of many collective phenomena observed in real-world populations, from the spread of behaviours such as smoking to political polarization and cultural trends~\cite{banisch2019opinion, bikhchandani1992theory}. As Keynes remarked, “worldly wisdom teaches that it is better for reputation to fail conventionally than to succeed unconventionally”~\cite{keynes36}. This captures the subtle power of conformity and the reputational risks embedded in social decision-making.

In this paper, we investigate a model in which a population of agents occupies a binary opinion state and updates their choices in response to both peer influence and an external stochastic signal, taking inspiration from interactions-based perspective in socioeconomic theory~\cite{granovetter1978threshold, galam2008sociophysics, holyst2001social, castellano2009statistical, durlauf1999can, bouchaud2013crises} and voter models~\cite{majmudar2020voter, jkedrzejewski2018impact, carro2016noisy}. We also acknowledge foundational work on ergodicity breaking in individual decision-making~\cite{meder2021ergodicity}, which motivates our investigation into the emergence of non-ergodic collective states. Specifically, our model aims to capture the conditions under which a population reaches a collective consensus or becomes trapped in a persistent state of agreement/disagreement with respect to a ground truth. These dynamics are closely related to findings in statistical physics, where binary systems and Ising-like models have been used to study phase transitions, symmetry breaking and collective behaviours~\cite{yeomans1992statistical, galam2004spontaneous, sood2005voter, mobilia2003does}. We incorporate stochastic influences to better reflect the unpredictability of social environments and the diversity of individual dispositions within real populations~\cite{brock2001discrete,lux1999scaling, kirman1992whom}. 

By situating our investigation at the intersection of opinion dynamics, game theory and statistical mechanics, we contribute to a growing body of work that explores how micro-level interactions and stochastic influences generate macro-level patterns in belief formation and consensus~\cite{nowak2010evolutionary, holme2006nonequilibrium, sornette2014physics, baronchelli2018emergence, gleeson2013binary}. Adopting this framework, we recover classic patterns of collective behaviour and reveal new factors that influence whether populations converge on truth or falsehood.

The manuscript is organised as follows. In Section~\ref{section2}, we describe the model and its dynamics, presenting both the memory-less and memory-based implementations. Section~\ref{section3} focuses on the high-sensitivity regime, where agents are more responsive to payoff differences and randomness plays a minor role. This setting allows for analytical insights. In Section~\ref{section4}, we turn to the low-sensitivity regime, where decision-making is more stochastic due to weaker agent sensitivity to payoff differences. In Section~\ref{sec:empirical}, we explore how the model can be empirically calibrated and applied to real-world phenomena such as voting and collective behaviour. Finally, Section~\ref{section5} presents our conclusions and outlines directions for future work. Supplementary results and analytical derivations are provided in the Appendix~\ref{sec:appendix}.

\section{Model implementation}\label{section2}
We investigate the evolution of a population of agents who can occupy a binary
state~\cite{granovetter1978threshold, galam2008sociophysics, galam2004unifying} and whose decisions are influenced by both peer behaviour and an external stochastic signal~\cite{holyst2001social, castellano2009statistical}. The dynamics are primarily governed by payoffs, which determine the agents’ probabilities of adopting or retaining a particular state~\cite{de2006statistical, banisch2019opinion}.

More precisely, at $t = 0$, all $N$ agents are initialised in the same state, either $1$ or $0$. These states correspond to binary choices, such as voting in a two-party system or adopting a social behaviour, e.g. smoking or not smoking.

At each time step $t = 0, \dots, T$, agents decide which state to adopt next on the basis of two factors. The first captures peer influence via $f_+(t)$, namely the fraction of agents in state 1 at time $t$, with $f_+(t) \in [0,1]$. When $f_+(t)\rightarrow0$ or $f_+(t)\rightarrow1$, consensus has been reached in the population. In the following, our analysis focuses on the behaviour of $f_+(t)$ without loss of generality, as $f_-(t)$ is implicitly defined by the normalization condition $f_-(t) = 1 - f_+(t)$.

The second component is an external signal $X(t)$, which takes value $+1$ with probability $p$ and $-1$ with probability $1 - p$. 
The payoff associated with each state is a convex combination of these components and is defined as:
\begin{equation}\label{eq:payoffs_memoryless_pp}
    \pi_\pm (t) = \gamma \, f_\pm(t) \pm (1 - \gamma)\, X(t),
\end{equation}
where the $+$ sign corresponds to state 1 and the $-$ sign to the opposite state. Here, $\gamma \in [0, 1]$ weighs the influence of peer behaviour: when $\gamma \rightarrow 1$, agents rely primarily on social conformity, while for $\gamma \rightarrow 0$ the external component dominates. When $X = 1$, agents are encouraged towards state $1$ as $\pi_+ > \pi_-$, whereas, when $X = -1$, they are driven towards state $0$ as $\pi_+ < \pi_-$. The negative sign in $\pi_-$ appropriately inverts the signal’s effect when considering $X = -1$. Note that, for the sake of conciseness, temporal dependence is omitted in the following when not necessary. 

The probability that an agent adopts state $1$ at the next time step is given by a Boltzmann distribution, as typical in decision-making models~\cite{brock2001discrete, weidlich2006sociodynamics}:
\begin{equation}\label{probability_of_changing}
    P_+(t) = \frac{e^{\beta \, \pi_+(t)}}{e^{\beta \, \pi_+(t)} + e^{\beta \, \pi_-(t)}},
\end{equation}
where the parameter $\beta$ plays a dual role as both the inverse temperature and a measure of the agents' sensitivity to payoff differences. As the inverse temperature, $\beta$ controls the level of randomness in decision-making: a high $\beta$ (low temperature) leads to more deterministic choices, where agents strongly prefer higher-payoff options, while a low $\beta$ (high temperature) introduces more randomness, making agents less responsive to payoff differences. From a behavioural perspective, $\beta$ can be interpreted as an intensity of choice parameter, quantifying how sharply agents react to variations in payoffs. Higher $\beta$ values indicate that agents are highly sensitive to even small differences in expected rewards, whereas lower $\beta$ values reflect a more indifferent, probabilistic decision-making process where payoff differences exert only a weak influence on the dynamics. Naturally, $P_- = 1- P_+$.
\begin{figure}[t]
    \centering
    \includegraphics[width=0.8\linewidth]{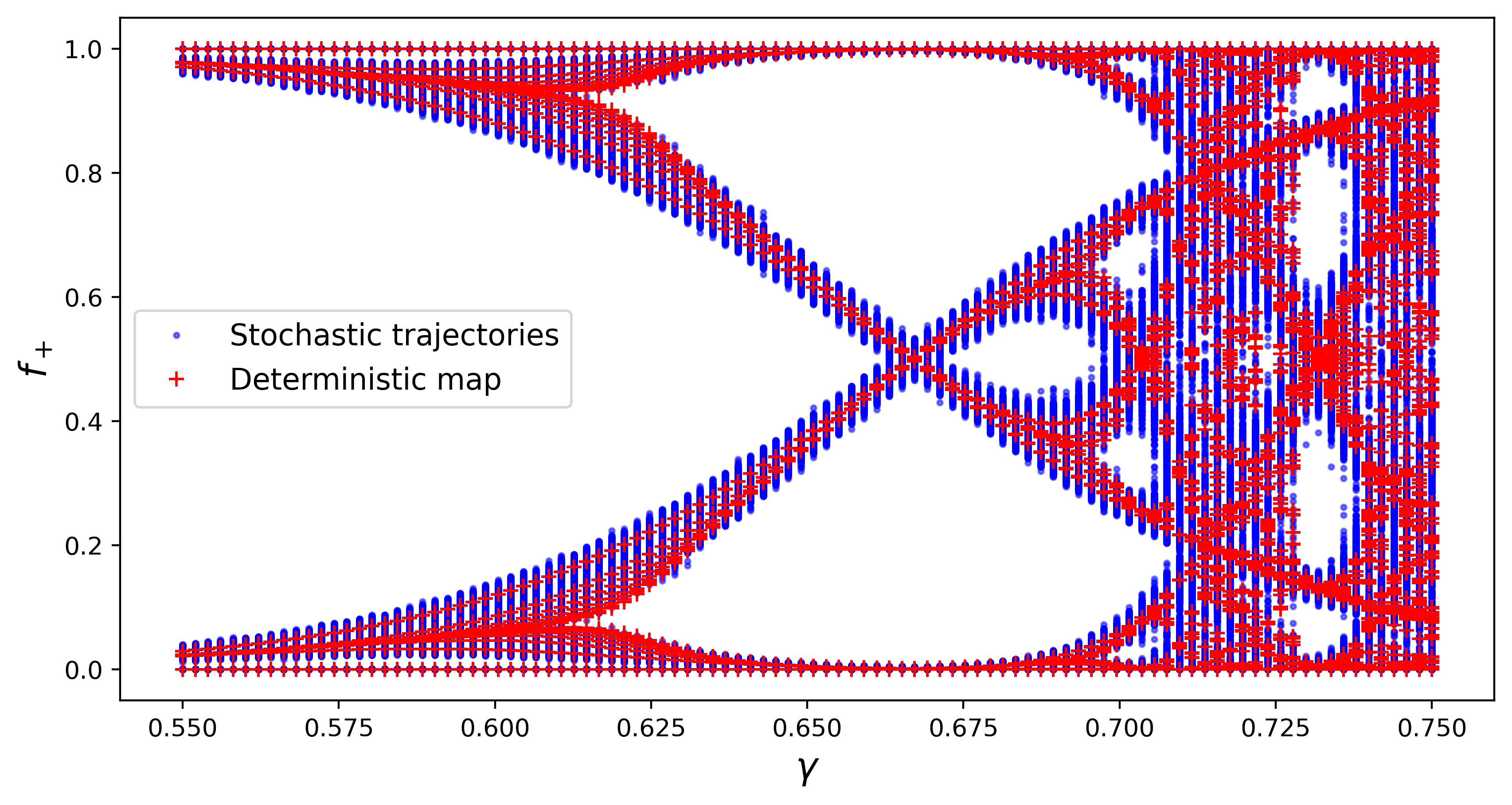}  
    \caption{Trajectories for stochastic and deterministic scenarios. Parameters: $\beta = 10, \, p=0.6,\,  N = 1000, \,T = 1000$. There are 15 iterations for deterministic map and the same number of independent simulations for stochastic dynamics.}
    \label{fig:trajectories}
\end{figure}
Given this implementation, our model offers a nuanced framework for decision-making, blending rationality (favouring higher payoffs) with randomness (occasionally choosing lower payoffs), reflecting real-world scenarios in which agents generally prefer optimal strategies but may still adopt suboptimal ones under uncertainty. This behaviour is characteristic of Fermi-like models, which originate in statistical physics~\cite{szabo1998evolutionary} and are widely used in evolutionary game theory~\cite{liu2015fixation}, social dynamics~\cite{castellano2009statistical, masuda2009evolutionary} and financial markets~\cite{chakraborti2009econophysics}.

In the aforementioned dynamics, payoffs depend on the instantaneous external signal, leading Eq.~\eqref{eq:payoffs_memoryless_pp} to describe a memory-less case. In contrast, we can add a memory-based framework that incorporates a finite memory, filtering out short-term fluctuations and making agents respond to a temporally smoothed version of the signal~\cite{moran2015post}. In this scenario, for $t \geq m$ agents compute the average of the last $m$ signal values:
\begin{equation}\label{eq:payoffs_memorybased_pp}
    \pi_\pm(t) = \gamma \, f_\pm(t) \pm (1 - \gamma)\, \langle X_m(t) \rangle,
\end{equation}
where
\begin{equation}\label{eq:def_xm}
    \langle X_m(t) \rangle = \frac{1}{m} \sum_{t' = t - m+1}^{t} X(t') \qquad \text{if} \quad t \geq m.
\end{equation}

In the following analyses, $\beta$ is held fixed to emphasise the effect of varying $\gamma$ across different scenarios, in line with the standard interpretation of $\beta$ in both physical and socio-economic contexts. Our focus is on the quantity $f_+$ across multiple simulations $S$, with the aim of identifying significant behavioural changes. We begin with the limit $\beta \rightarrow \infty$, as it exhibits the most characteristic features and dynamics that are suitable to analytical treatment. We then proceed to lower values of $\beta$. 

Unless otherwise specified in the figure captions, simulations are performed with $N = 500$, $p = 0.6$, and $T = 1000$, which represent standard settings used throughout the paper. The number of agents, $N$, is chosen to balance computational feasibility with sufficient averaging over population-level effects. The signal bias $p = 0.6$ reflects a mildly asymmetric environment ($p > 0.5$), allowing clearer observation of coordination effects and asymmetries in the collective response. The number of time steps $T$ ensures that the system reaches a stationary regime. The sample size $S$ (i.e., number of independent realizations) is selected to provide reliable ensemble averages while keeping computational cost manageable. In the memory-based setting, we fix the memory length to $m = 200$, which is large enough to induce smoothing of short-term fluctuations without completely suppressing responsiveness to changes in the external signal. A key feature of the model is its symmetry under the transformation $p \rightarrow 1 - p$. Although $p$ formally spans $[0, 1]$, we focus on $p \in [0.5, 1]$, as flipping $p$ is equivalent to inverting the signal ($X \rightarrow -X$), which simply exchanges the roles of the two states. Consequently, the system remains invariant under a reversal of the signal and relabelling of the states.

By examining the trajectories of $f_+$ over time across multiple simulations, we obtain Fig.~\ref{fig:trajectories}, which displays both the stochastic dynamics and their deterministic counterpart. Here, $f_+$ denotes the fraction of agents in state $+1$, and its evolution reflects the collective opinion of the system. The stochastic trajectories (blue markers) are generated by simulating the agent-based dynamics for several values of $\gamma$. At every time step, agents observe the current population state and a randomly drawn external signal $X(t) \in \{-1, +1\}$, with probability $p$ of being $+1$. Using this input, each agent updates their state probabilistically, following a Fermi-type rule determined by the payoffs. This process is repeated for $T$ time steps, and the full time series of $f_+(t)$ is recorded. The figure shows, for each $\gamma$, the complete set of $f_+$ values visited during the simulations, across time and across 15 independent realizations,  plotted along the vertical axis for each $\gamma$.

In contrast, the deterministic counterparts (red markers) are obtained by iterating a noiseless update map based on the model equations. Starting from two initial conditions, $f_+ = 0$ and $f_+ = 1$, we recursively apply the update rule. At each step, we consider both possible values of the external signal, $X = \pm 1$,  effectively generating a binary tree of possible $f_+$ evolutions. After 15 iterations, we collect the set of unique $f_+$ values that emerge, which are dynamically accessible under deterministic dynamics. 

This representation reveals that the behaviour of the agents (when averaged over a large ensemble and long time horizons) closely resembles the deterministic dynamics.  It also highlights the system’s diverse behaviours depending on the value of $\gamma$: for low $\gamma$, the system tends toward extremal states ($f_+ \approx 0$ or $f_+ \approx 1$), while for higher $\gamma$, the trajectories become increasingly complex and fragmented. However, this representation is incomplete: it indicates which values of $f_+$ are visited but offers no insight into their temporal persistence, stability, or ergodic properties. Therefore, in what follows, we turn to alternative representations that more effectively capture the system’s underlying mechanisms and dynamic behaviour. In particular, in the memory-based setting, agents evaluate actions based on the recent history of the external signal, rather than on its instantaneous value alone. This introduces temporal inertia, reducing sensitivity to noise and fostering long-term adaptation to persistent environmental trends.

\section{High-Sensitivity Regime (Low Noise Limit)}\label{section3}
Our analysis begins in the limit $\beta \to \infty$, corresponding to a high-sensitivity (low-noise) regime where agents respond strongly to payoff differences. For practical purposes, this regime is approximated in simulations by setting $\beta = 100$. A sensitivity analysis confirms that the qualitative behaviour remains robust for other large values of $\beta$, such as $\beta = 50$ or $\beta = 20$.
\begin{figure}[t]
    \centering
    \includegraphics[width=0.75\linewidth]{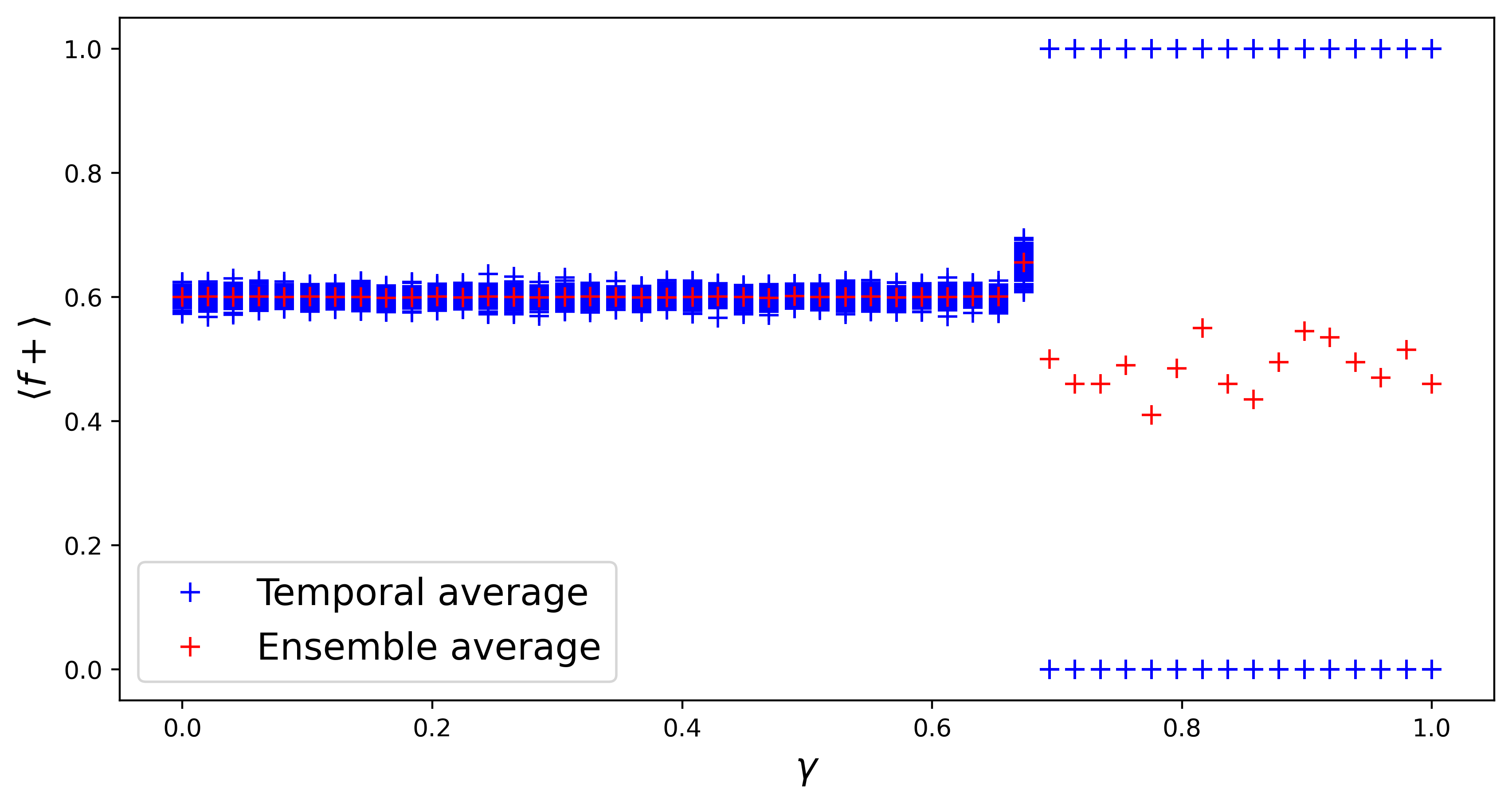} \, 
    \includegraphics[width=0.75\linewidth]{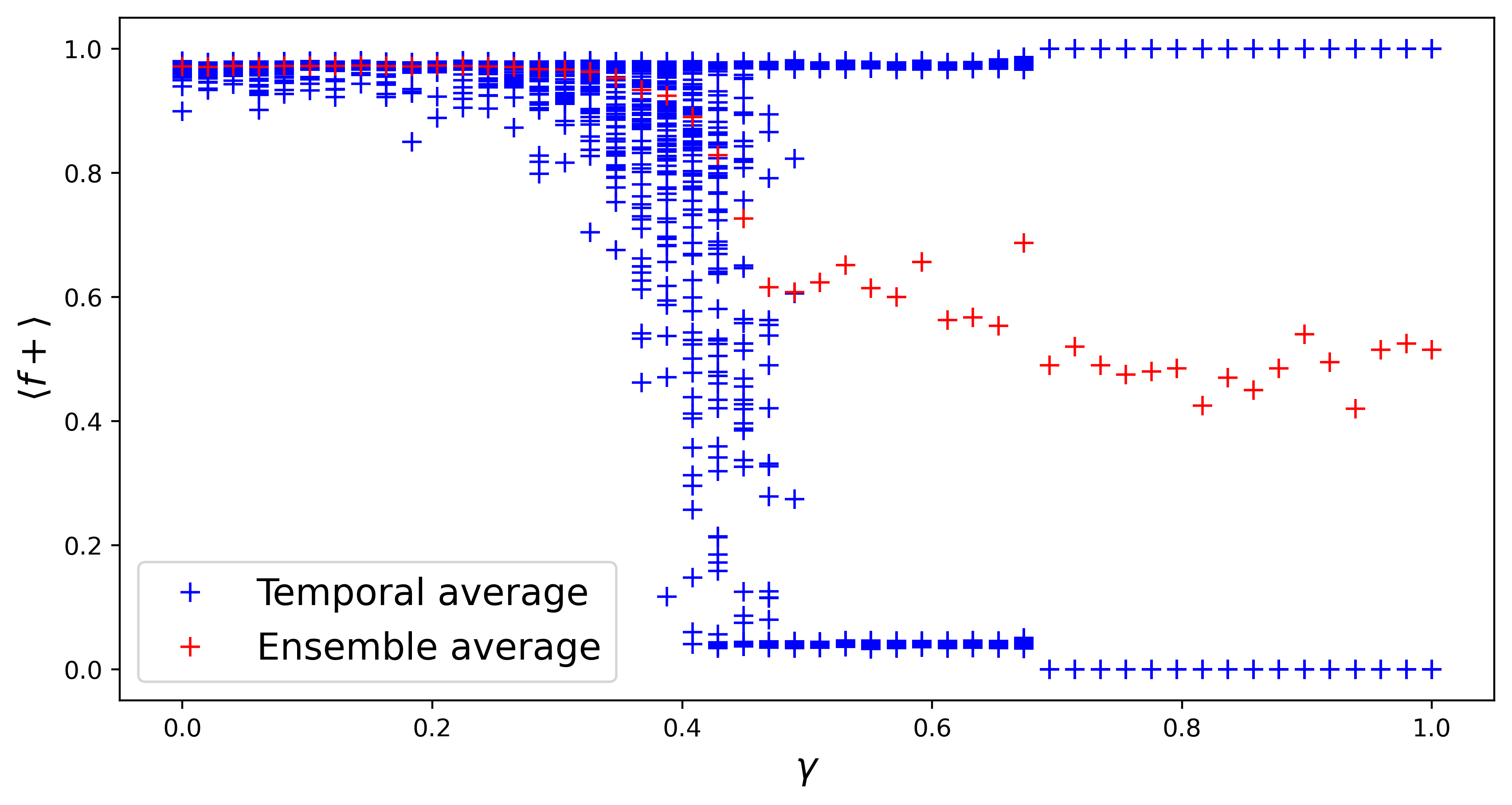} 
    \caption{Temporal and ensemble averages across several simulations. Simulation parameters are: $\beta = 100, \, p=0.6,\, N = 500$, number of time steps $\,T = 3000$ and sample size $S = 200$. Upper panel: Memory-less scenario, with $\gamma^* = 0.69$. Lower panel: Memory-based scenario, averaging the last $m=200$ values of the stochastic signals, with $\gamma^* = 0.37$.}
    \label{fig:mean_beta100}
\end{figure}

Fig.~\ref{fig:mean_beta100} shows the results of multiple independent simulations for varying values of $\gamma$, the social conformity parameter. In each simulation, the temporal mean of the observable $f_+$ is computed within each simulation (blue markers, each indicating a single realization) and is compared with the ensemble average, i.e., the mean of the temporal averages across all simulations (red marker). When these two quantities (denoted as $\langle f_+\rangle $) coincide, the system is said to be ergodic: time averages are representative of ensemble statistics and the dynamics are self-averaging. We refer to this regime as the oscillating phase, in which agents fluctuate between state 0 and state 1 (as confirmed by Figs.~\ref{fig:trajectories} and~\ref{fig:jumps}), spending approximately a fraction $p$ of the time in state $+1$, yielding 
\begin{equation*}
    \langle f_+\rangle \simeq p.
\end{equation*}
This behaviour persists up to a critical value of $\gamma$, denoted $\gamma^*$. Beyond this point, a qualitative change occurs and agents enter the so-called absorbing phase: temporal averages become widely scattered and their correspondence with the ensemble average deteriorates. This divergence reflects a breakdown of ergodicity, as agents tend to become trapped in a single state for the remainder of the simulation. Fig.~\ref{fig:mean_beta100} displays both the memory-less (upper panel) and memory-based (lower panel) scenarios, clearly highlighting their distinct behaviours. Incorporating memory leads to a less abrupt transition, with the onset of the absorbing phase occurring at significantly lower values of $\gamma$ compared to the memory-less case. After the transition, in both cases, the ensemble average stabilises around 
\begin{equation*}
    \langle f_+\rangle \simeq 0.5.
\end{equation*}
Notably, to identify the value of $\gamma^*$ in the memory-based scenario, we consider the smallest value of $\gamma$ at which a noticeable discrepancy arises between the temporal means and the overall average.

\subsection{Focus on memory-based dynamics}
In the high-sensitivity regime, some analytical insights can be gained. Specifically, in the memory-based scenario, a ``mean-field approximation" can be employed. Thus, the time-averaged signal $\langle X_m(t)\rangle$ is replaced by its average, yielding the limiting equation
\begin{equation}
    \langle X_m(t)\rangle \xrightarrow{m\rightarrow\infty} 2 \, p -1,
\end{equation}
which allows Eq.~\eqref{eq:payoffs_memorybased_pp} to be rewritten as
\begin{equation}\label{eq:payoffs_memoryless_pp_p}
    \pi_\pm(t) = \gamma \, f_\pm(t) \pm (1-\gamma)\, (2 \, p -1),
\end{equation}
This leads to an effective deterministic description of the system's average behaviour.

\begin{figure}
    \centering
    \includegraphics[width=0.8\linewidth]{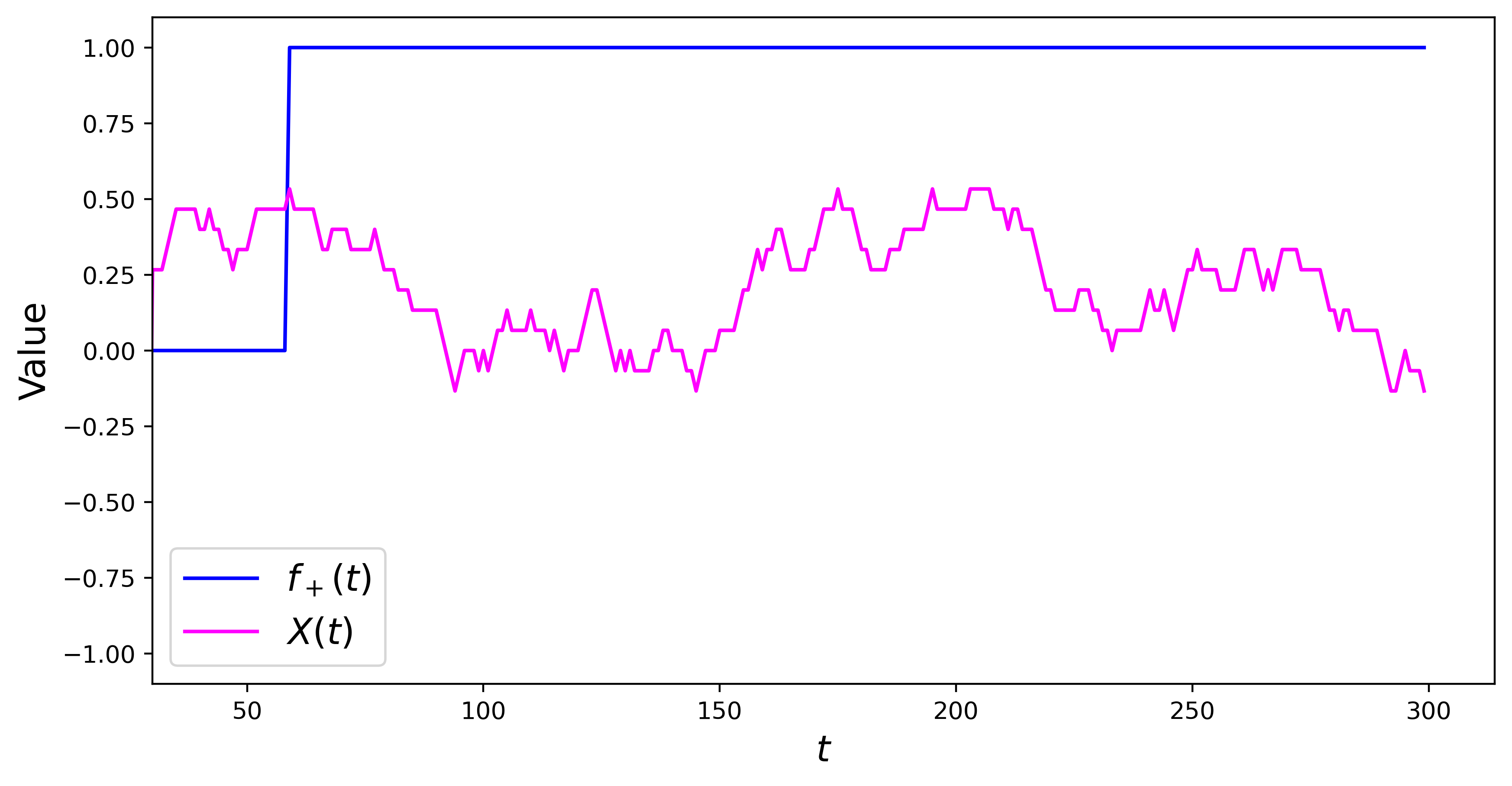}  
    
    \caption{The blue line shows the trajectory of a single agent, clearly illustrating a  the external signal $X(t)$, shown in pink, exceeds a certain threshold. Up to time $t = m-1$, $X(t)$ corresponds to its instantaneous value; from $t = m$ onward, it is defined as the average of the last $m$ values. Parameter values are: $\beta = 100, \, p=0.6,\,  N = 1, \,T = 300, \,m = 30, \, \gamma = 0.5$. }
    \label{fig:jumps}
\end{figure}
When the system evolves over a sufficiently long time (i.e., $T > 100$) and long memory is considered (specifically, $m > 20$), it is possible for a sudden flip to occur from state 0 to state 1, after which the agent becomes trapped in state 1. This behaviour is illustrated in Fig.~\ref{fig:jumps} for a single agent. We investigate the conditions under which such flips can occur. By inserting $f_+ = 0$ into Eq.~\eqref{eq:payoffs_memorybased_pp} and equating the payoffs of the two states, we obtain:
\begin{equation}
    (1-\gamma ) \langle X_m \rangle  = \gamma - (1- \gamma) \langle X_m \rangle.
\end{equation}
Solving for $\langle X_m \rangle$, the result is
\begin{equation}\label{eq:x_star}
    X^* = \frac{\gamma}{2\, (1-\gamma)},
\end{equation}
where $X^*$ represents the threshold value of $\langle X_m \rangle$ required for a flip from state 0 to state 1 to occur. Note that, as $\gamma$ increases, $X^*$ also increases, eventually reaching values for which no feasible solutions exist. Importantly, this relation is independent of $p$.
\begin{figure}
    \centering
    \includegraphics[width=0.8\linewidth]{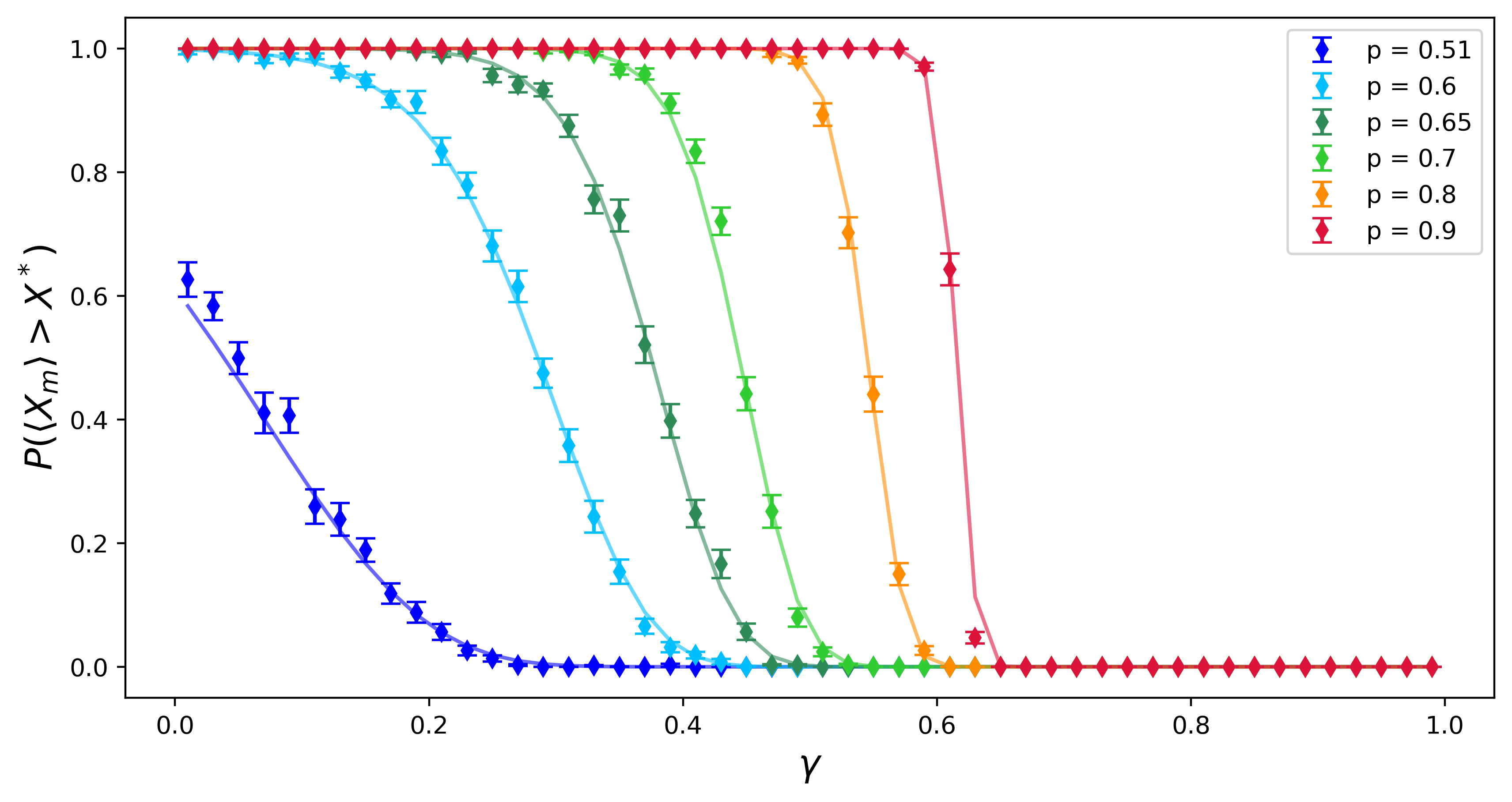}  
    \caption{Probabilities of flipping $P\left (\langle X_m\rangle\right ) > X^*)$ from state 0 to state 1 for several values of $p$, with a comparison of analytical (lines) and computational (markers with error bars). Other parameters are: $\beta = 100$, $N = 100$, $T = 1000$, $m = 200, S = 50$.}
    \label{fig:probflips}
\end{figure}

Within this framework, we study the probability of exceeding the threshold $X^*$. We adopt both computational and analytical approaches, focusing on the probability that the empirical average $\langle X_m \rangle$ exceeds $X^*$, namely:
\begin{equation}\label{eq:probabilityofflip}
    P\left (\frac{1}{m} \sum_{i =1}^{m} X(i) > X^*\right).
\end{equation}
The analytical treatment is presented in Appendix~\ref{sec:appendix}, yielding the result:
\begin{equation}\label{eq:probabilityth}
    P\left( \frac{1}{m} \sum_{i=1}^m X(i) > X^* \right)
\approx \frac{1}{2} \left[ 1 - \operatorname{erf}\left( \frac{\sqrt{m}\, (1 + X^* - 2p)}{2\sqrt{2\,p\, (1 - p)}} \right) \right],
\end{equation}
where $m$ is the number of time steps considered in computing the mean. In Fig.~\ref{fig:probflips} we compare the analytical predictions (lines) and numerical simulations (markers with error bars), finding strong agreement, concluding that our stylized model can correctly reproduce theoretical features.  In the simulations, we track how often the condition in~\eqref{eq:probabilityofflip} is satisfied in the memory-based scenarios, namely when $\langle X_m\rangle$ exceeds the critical threshold defined in Eq.~\eqref{eq:x_star}. The crossing frequency is recorded per simulation and averaged across several independent realizations $S$ to obtain the empirical flipping probability. This value is then compared to the theoretical probability predicted from Gaussian statistics using the error function. 

For each fixed value of the signal bias $p$, the flipping probability remains close to 1 at low values of the social conformity parameter $\gamma$, then rapidly drops to 0 near a critical value. This behaviour reflects the increasing difficulty of exceeding the threshold $X^*$ given by Eq.~\eqref{eq:x_star}, which grows with $\gamma$. For fixed $\gamma$, a higher $p$ increases the probability of transition to state 1.

\subsection{Critical Weight of Conformity}
To further investigate the transition from oscillating to absorbing behaviour, we analyse the fixed points of the mean-field dynamical system emerging from the memory-based dynamics and assess their stability. Specifically, we start from the dynamics governed by Eq.~\eqref{eq:payoffs_memorybased_pp}, with a memory-based and “mean-field approximation”, which means that the payoff is given by
\begin{equation}
    \pi_\pm(t) = \gamma\, f_\pm(t)\pm(1-\gamma) \,(2\,p-1). 
\end{equation}
From here, we derive a self-consistent equation for the expected fraction $f_+(t+1)$ of agents in the state +1 at the next time step, given the current fraction $f_+(t)$:
\begin{equation}
f_+(t+1) = \frac{1}{1 + \exp\left(-\beta \left[ 2\gamma f_+(t) + 2(1-\gamma)(2p - 1) - \gamma \right] \right)}.
\end{equation}

The fixed points of the system correspond to the values of $f_+$ that satisfy
\begin{equation*}
    f_+(t+1) = f_+(t),
\end{equation*}
which can be identified numerically. Their stability is determined by evaluating the derivative $f_+'(t+1)$, as a fixed point is considered stable if small perturbations decay over time, i.e., if
\begin{equation}
    |f_+'(t+1)| < 1.
\end{equation}
\begin{figure}
    \centering
    \includegraphics[width=0.8\linewidth]{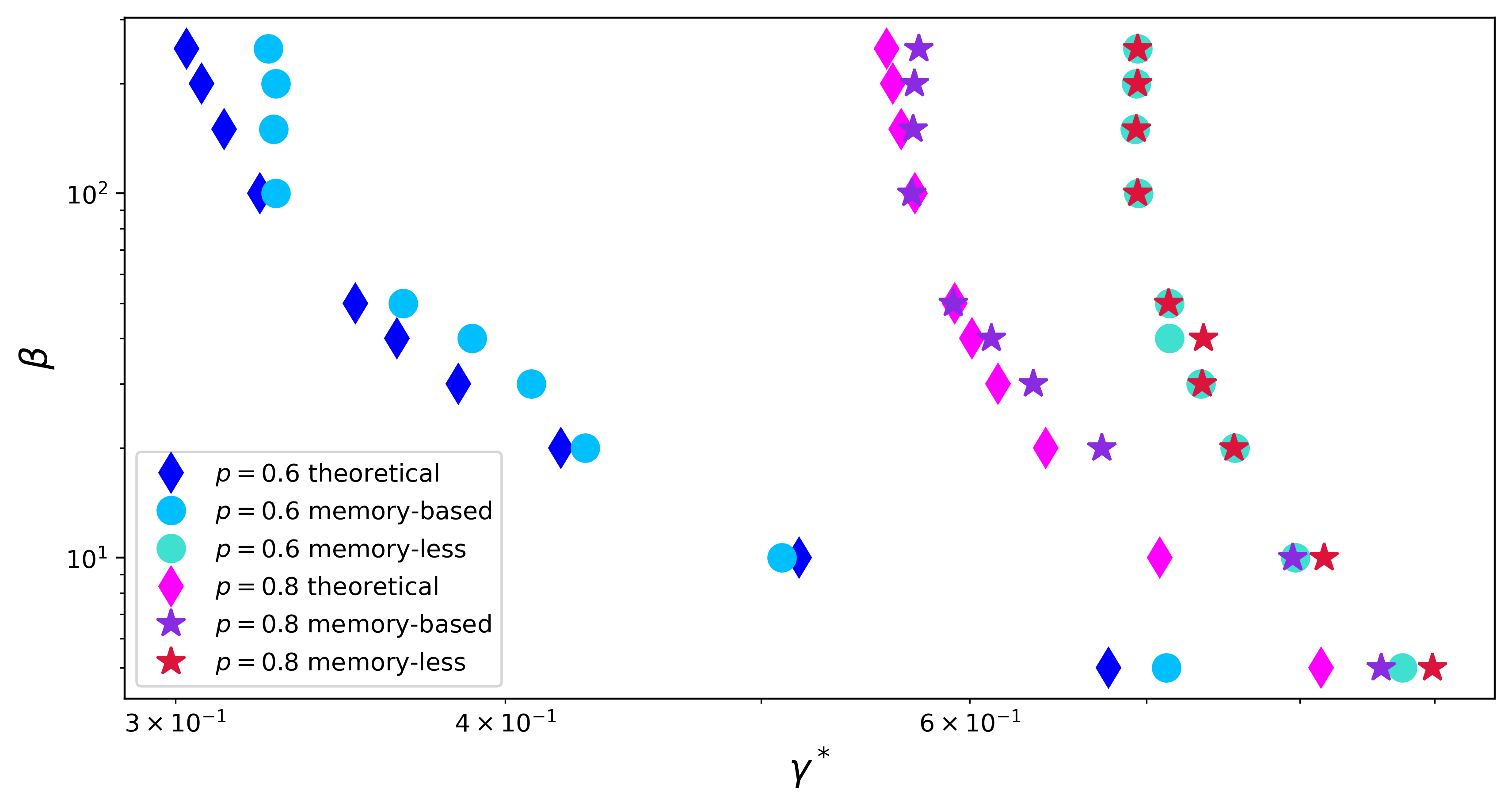}
    \caption{Comparisons in log-log scale of the critical value $\gamma^*$ for theoretical, memory-based and memory-less implementations, setting $p=0.6, \,0.8$. For the latter two, parameters are: $ N = 100,\, T = 1000,\, S = 50, \,m = 200.$}
    \label{fig:phasediagram}
\end{figure}
By varying $\gamma$ while keeping other parameters fixed, this analysis enables the identification of the critical value $\gamma^*$ at which a bifurcation occurs, corresponding to the emergence of a third, unstable fixed point between two stable ones.

These results are compared with the critical values of $\gamma^*$ observed in simulations for the memory-based and memory-less scenarios. The resulting phase diagram~\cite{holme2006nonequilibrium}, shown in Fig.~\ref{fig:phasediagram}, illustrates the transition: for $\gamma < \gamma^*$ (below the transition curve), the system exhibits oscillatory dynamics, whereas for $ \gamma > \gamma^*$ (above the curve), it enters the absorbing phase. Notably, the memory-based scenarios align more closely with the theoretical predictions compared to the memory-less cases, as it is closer to the limit behaviour.  While small discrepancies may arise due to finite-size effects, the mean-field approach nonetheless provides valuable qualitative insights into the system’s behaviour, particularly in capturing phase transitions and approximating fixed points. 
\subsection{Networks}
\begin{figure}
    \centering[t]
    \includegraphics[width=0.75\linewidth]{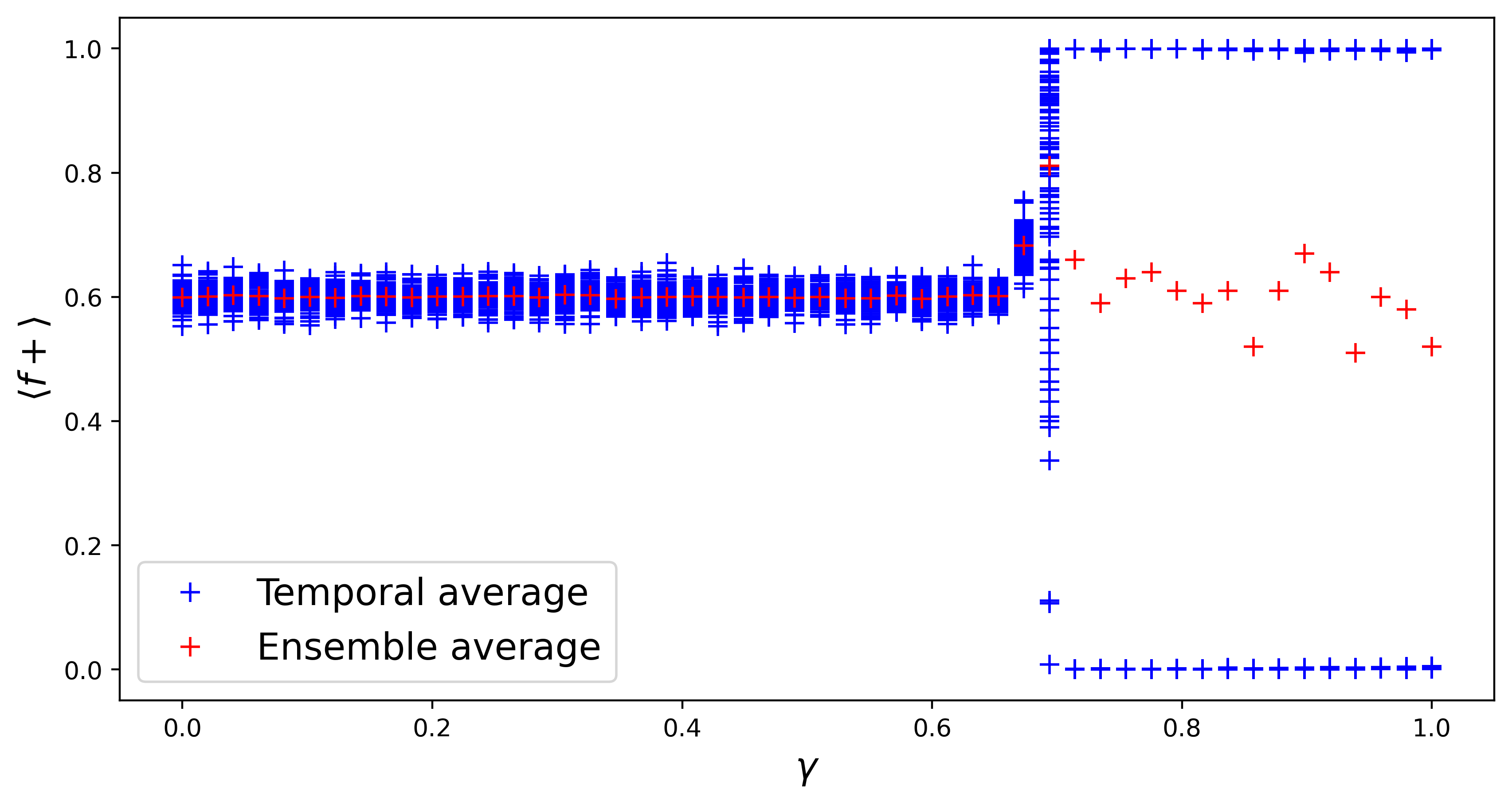}  
    
    \includegraphics[width=0.75\linewidth]{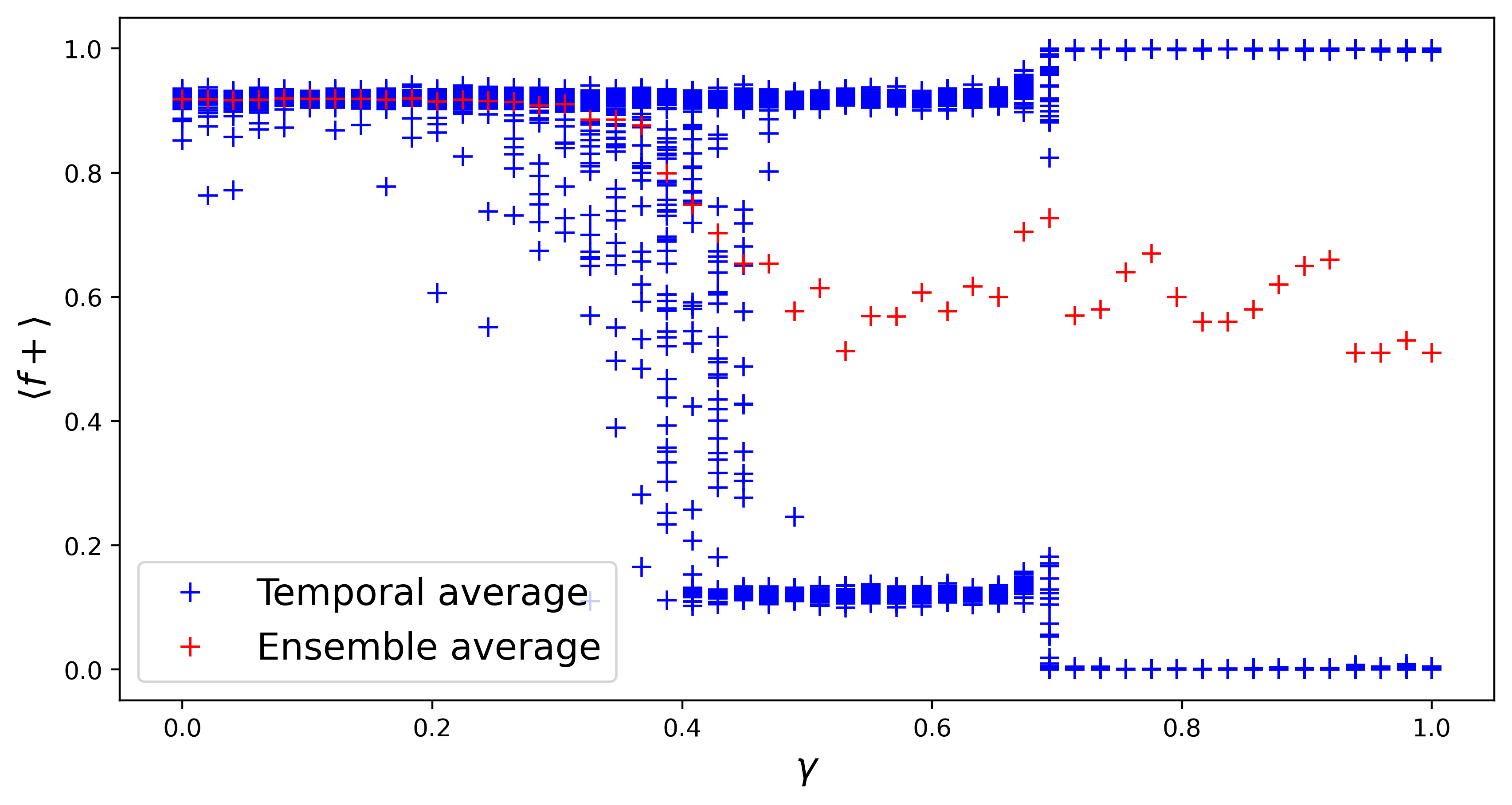}
    \caption{Temporal mean and ensemble average for Erd\H{o}s–Rényi implementation. Parameters: $T= 1000, \, p =0.6$, sample size $S = 100$, average degree $\langle k \rangle =10$. Upper panel: Memory-less scenario, with $\gamma^* = 0.71$. Lower panel: Memory-based scenario, averaging the last $m=200$ values of the stochastic signals, with $\gamma^* = 0.33$.}
    \label{fig:networks}
\end{figure}
The obtained results can be extended to cases where the dynamics is implemented on a network. We first analyse the Erd\H{o}s–Rényi network~\cite{erdos1960evolution}, which approximates the well-mixed population scenario~\cite{evdynamics}. When applied to stochastic block models~\cite{holland1983stochastic} (networks with modular community structure) we observe the same qualitative behaviour as in the Erd\H{o}s–Rényi case, provided that two or more communities are present. We then extend our analysis to the Barabási–Albert model~\cite{barabasi1999emergence}, a scale-free network with heterogeneous degree distributions arising from preferential attachment.

Across all three network types, we observe the same distinctive behaviour: an oscillatory phase for values of the social influence parameter $\gamma < \gamma^*$ and an absorbing phase for $\gamma > \gamma^*$, with a critical value $\gamma^* \simeq 0.7$ in the memory-less case and a lower value in the memory-based one. Remarkably, the phase transition occurs at the same critical point as in the well-mixed scenario, indicating that the emergence of this sharp transition is robust to the underlying network topology. Our findings align with those of previous studies exploring noisy voter models on networks~\cite{carro2016noisy}. In addition, our focus on the onset of ergodicity breaking across topologies introduces a further analytical layer, allowing for comparisons between memory-based and memory-less regimes.

Results for the  Erd\H{o}s–Rényi are shown in Fig.~\ref{fig:networks} as representative of the network-based scenarios. 

\section{Low sensitivity}\label{section4}
We now turn our attention to the low-sensitivity regime, defined by $\beta \in [0,1]$, where $\beta$ represents the agents’ sensitivity to payoff differences (or, equivalently, the inverse temperature in the decision rule). Within this range, agents behaviour becomes increasingly random, approaching full stochasticity in the limit $\beta \to 0$~\cite{galam2008sociophysics, galam2011sociophysics}.
\begin{figure}[t]
    \centering
    \includegraphics[width=0.8\linewidth]{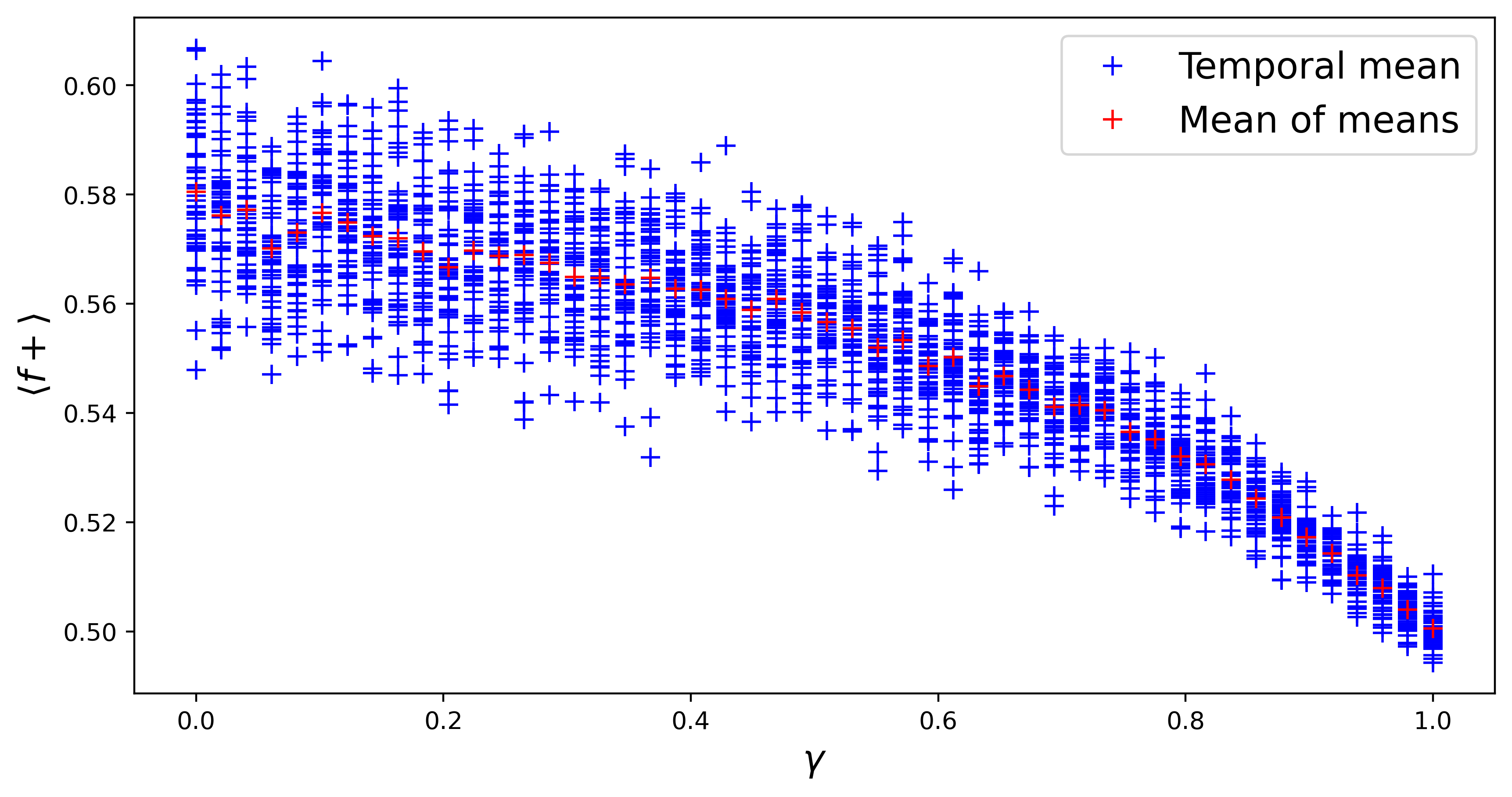} \, 
    \caption{Temporal mean and ensemble average across several simulations, for memory-less scenario. Parameters: $\beta = 1,\, p=0.6, \, N = 100, \, T = 1000,\, S = 50.$}
    \label{fig:mean_beta1}
\end{figure}

Starting from $\beta = 1$, no critical transition is observed, but a decreasing trend in $\langle f_+ \rangle$ emerges, as shown in Fig.~\ref{fig:mean_beta1}. When $\gamma$ is small, the external signal primarily drives the dynamics, yet its effect is dampened by the low value of $\beta$, leading to the approximation
\begin{equation*}
\langle f_+ \rangle \sim p.    
\end{equation*}
Notably, a large spread in temporal means is observed due to the stochasticity of individual realisations. Conversely, for large values of $\gamma$, peer influence dominates, neutralising the bias imposed by the external signal and resulting in
\begin{equation*}
    \langle f_+ \rangle \sim 0.5.
\end{equation*}
Thus, strong social conformity can neutralise the effect of external persuasion, even when the external information is systematically biased. Moreover, the spread in temporal means narrows for large $\gamma$, reflecting increased alignment among agents and reduced variability. Memory-less and memory-based dynamics yield similar outcomes in this regime, as the memory mechanism merely smooths the signal without altering the dominant behaviour.

In the extreme case $\beta = 0$, agents' decisions become entirely random: each state is adopted with equal probability $ p = 0.5$, independent of both peer influence and the external signal. This corresponds to an infinite-temperature regime, where the energy landscape is completely flat and no directional bias in the dynamics exists. As a result, the fraction of agents in state $+1$ fluctuates around the unbiased value $\langle f_+ \rangle = 0.5$.

As in the high-sensitivity regime, similar results are obtained when the dynamics are implemented on networks.

\section{Applications and Empirical Relevance}\label{sec:empirical}
Our model lends itself to practical applications in empirical contexts, particularly in understanding real-world collective phenomena such as voting behaviour, public opinion formation, and social mobilization. Central to such applications is the possibility of estimating or calibrating the critical threshold $\gamma^*$, which demarcates the transition from an ergodic regime (characterized by flexibility and responsiveness to external stimuli) to a non-ergodic regime (marked by persistent consensus or polarization).

A first methodology for empirically anchoring the model involves survey-based calibration, where data from social and political surveys (e.g. Eurobarometer and the General Social Survey,) can be used to quantify the relative importance individuals assign to peer influence versus external information. This enables estimation of the conformity weight $\gamma$ at a population level~\cite{banisch2019opinion, bikhchandani1992theory}.

A second approach relies on behavioural experiments, in which participants make binary decisions (e.g., supporting or opposing a policy) under controlled exposure to both peer choices and external cues such as expert recommendations~\cite{centola2010spread}. In such settings, the critical value $\gamma^*$ can be inferred by analysing shifts in decision patterns. These designs allow fine-grained estimation of how social influence and external signals jointly shape individual and collective decision-making.

A third empirical path involves data-driven modelling of online behaviour, especially in digital environments where opinion dynamics unfold over time. Platforms such as Twitter, Reddit, or online petition sites provide natural environments in which users observe both social consensus (e.g., likes, retweets) and external signals (e.g., news or expert commentary).The emergence and reversibility of large-scale opinion cascades may indicate proximity to the critical point $\gamma^*$. Previous studies have shown that social imitation can drive abrupt shifts in collective opinions, consistent with ergodicity-breaking transitions~\cite{michard2005theory,lorenz2011social, moussaid2013social}.

These empirical tools can be applied to specific real-world contexts. In voting behaviour, for instance, public opinion data preceding major referenda or elections (such as the Brexit vote or U.S. presidential elections) often display early fluidity followed by sharp polarization~\cite{baldassarri2007dynamics, hobolt2016brexit}. Our model predicts such dynamics when the social weight $\gamma$ crosses a critical threshold. By analysing the variance in opinion trajectories across regions or demographic groups,  it is possible to infer whether different communities operate within ergodic or absorbing regimes~\cite{boxell2024cross}. In the context of social action and behaviour adoption, examples such as climate strike participation, vaccination campaigns, or online mobilization efforts also show sensitivity to social reinforcement. Experimental and observational studies have shown that reaching a critical mass of adopters can trigger self-sustaining cascades~\cite{centola2010spread}. This dynamic is captured in our model by the transition from oscillating to absorbing states. The model’s prediction that strong social conformity can dampen responsiveness to even biased external signals aligns with observed behavioural inertia in tightly networked or ideologically homogeneous populations. 

In summary, the model’s parameters are not merely theoretical constructs but are measurable and estimable through a variety of empirical methods. Calibrating the model to real-world data opens a pathway for using the critical threshold $\gamma^*$ as a predictive indicator of opinion stasis, entrenched consensus, or the emergence of sudden collective shifts. These connections provide a basis for applying the model to policy-relevant domains such as political campaigning, public health interventions, and media strategy. 

\section{Discussion}\label{section5}
We introduce a stochastic imitation-based model of opinion dynamics, in which agents balance social conformity with responsiveness to an external signal. The model operates in a binary setting, capturing a wide range of collective behaviours such as voting~\cite{granovetter1978threshold}, behavioural adoption~\cite{schelling1978micromotives}, or participation in social action~\cite{centola2010spread}. Our framework takes inspirations from existing models of noisy voter dynamics and memory effects~\cite{carro2016noisy, jkedrzejewski2018impact}, it departs from previous approaches by characterizing ergodicity breaking as a phase transition driven by the interplay of conformity, randomness, and external bias. Unlike jump process-based models~\cite{majmudar2020voter}, our model captures continuous decision probabilities and analytically derives the critical threshold for phase transitions, contributing a new perspective on collective adaptation and inertia.

In addition, beyond its theoretical relevance, the model also lends itself to empirical calibration and interpretation. In particular, the critical threshold of the social weight $\gamma^*$ may be estimated from survey data, behavioural experiments, or digital traces of social dynamics~\cite{centola2010spread}.

Our analysis reveals a critical value critical value of the social weight, $\gamma^*$, separating two distinct phases. For $\gamma < \gamma^*$, the system remains ergodic: agents fluctuate between states and the population correctly tracks the external signal. Beyond $\gamma^*$, the system becomes non-ergodic, with agents locking into persistent opinion states, reflecting a breakdown of adaptability~\cite{castellano2009statistical, galam2012sociophysics}. This transition is robust across both memory-less and memory-based dynamics, although memory smooths the transition and lowers the critical threshold~\cite{sornette2014physics}. While memory reduces noise and fosters stability, it can also delay adaptation to new information, echoing behavioural inertia observed in organisational and economic settings~\cite{simon1982models}.

The critical threshold $X^*$ that the average external signal must exceed for triggering collective shifts can be analytically characterised, enabling predictions about the likelihood of transitions. These results offer insight into real-world phenomena such as bubbles, viral trends or sudden changes in public sentiment~\cite{bouchaudforecasting, shiller2015irrational}. Network simulations confirm that the observed transitions are not artefacts of specific topologies: from Erdős–Rényi~\cite{erdos1960evolution} to scale-free~\cite{barabasi1999emergence} and modular networks~\cite{holland1983stochastic}, the phase transition persists. This robustness supports the model’s applicability across a range of empirical contexts (including political discourse, behavioural contagion, and financial markets) where network structures may vary significantly but imitation dynamics persist.

Our framework relates to empirical and theoretical work on imitation as a driver of tipping points in social systems~\cite{livan2019don,de2024imitation, peters2016evaluating}, including applause dynamics~\cite{michard2005theory}, prediction markets~\cite{wolfers2004prediction} and financial herding~\cite{cont2000herd}. Furthermore, our model connects with stochastic Ising-type extensions~\cite{brock2001discrete} and with frameworks such as the minority game~\cite{challet2005minority}, which may complement future investigations.

In practical terms, the model suggests that in highly conformist settings, targeted interventions (e.g., through network hubs) may be more effective than broadcasting external information~\cite{jackson2010social}. Conversely, in less conformist contexts, improving the quality and clarity of external signals may play a more significant role in shaping collective outcomes.

Future research could explore adaptive $\gamma$, the weight of social conformity, and $\beta$,  the agents’ sensitivity to payoff differences (or inverse temperature) values, as well as endogenous external signals~\cite{DeMartino_2006} or heterogeneous agent populations~\cite{kirman1992whom, lux1999scaling}. These extensions would bring the model closer to the complexity of real-world social systems and offer richer insight into collective decision-making under uncertainty.

\section*{Data availability statement}
No new data were created or analysed in this study.

\appendix
\section{Appendix: Analytical derivation of the critical flipping probability}\label{sec:appendix}
In the high-sensitive memory-based regime, we aim to study the probability that the empirical average $\langle X_m (t)\rangle$, defined in Eq.~\eqref{eq:def_xm}, exceeds the threshold value $X^*$, required for a flip from state 0 to state 1 to occur, as seen in Eq.~\eqref{eq:x_star}. This amounts to compute:
\begin{equation}
    P\left (\frac{1}{m} \sum_{i =1}^{m} X(i) > X^*\right).
\end{equation}
Defining $ S_m = \sum_{i=1}^m X(i)$, we observe that $S_m$ can be expressed in terms of a binomial random variable
\begin{equation*}
    S_m = 2Y - m,
\end{equation*}
where $Y \sim \mathrm{Bin}(m, p)$ counts the number of $X(i)$ equal to $ 1$. We can then rewrite the probability
\begin{equation*}
    P \left( \frac{S_m}{m} > X^* \right) = P\left( \frac{2Y - m}{m} > X^* \right) = P\left( Y > \frac{m(1 + X^*)}{2} \right).
\end{equation*}
Indicating $k^* = \frac{m(1 + X^*)}{2}$, we apply the normal approximation to the binomial distribution
\begin{equation*}
    Y \approx \mathcal{N}(mp, mp(1 - p)),
\end{equation*}
and then standardise
\begin{equation*}
P\left( Y > k^* \right) \approx P \left( Z > \frac{k^* - mp}{\sqrt{mp(1 - p)}} \right),
\end{equation*}
where $ Z \sim \mathcal{N}(0, 1)$. Substituting  $k^*$, we obtain
\begin{equation*}
\frac{k^* - mp}{\sqrt{mp(1 - p)}} = \frac{m(1 + X^* - 2p)}{2\sqrt{mp(1 - p)}}
= \frac{\sqrt{m}(1 + X^* - 2p)}{2\sqrt{p(1 - p)}}.
\end{equation*}
Hence, the probability becomes
\begin{equation*}
    P\left( \frac{1}{m} \sum_{i=1}^m X(i) > X^* \right)
\approx \frac{1}{2} \left[ 1 - \operatorname{erf}\left( \frac{\sqrt{m}(1 + X^* - 2p)}{2\sqrt{2p(1 - p)}} \right) \right],
\end{equation*}
where $ \operatorname{erf} $ denotes the Gauss error function. 

We can compute the limiting behaviour as $m\to\infty$. According to the Law of large numbers, we know that
\begin{equation*}
\frac{1}{m} \sum_{i=1}^m X(i) \xrightarrow{\text{a.s.}} \mathbb{E}[X(i)] = 2p - 1.    
\end{equation*}
Therefore,
\begin{equation*}
\lim_{m \to \infty} P\left( \frac{1}{m} \sum_{i=1}^m X(i) > X^* \right) =
\begin{cases}
1, & \text{if } X^* < 2p - 1, \\
\frac{1}{2}, & \text{if } X^* = 2p - 1, \\
0, & \text{if } X^* > 2p - 1.
\end{cases}    
\end{equation*}

This reflects the fact that the empirical mean converges to its expected value $2p - 1$, and the probability of deviating from this value becomes negligible as $m$ grows large. The excellent agreement between computational and analytical results in Fig.~\ref{fig:probflips} supports the validity of the Gaussian approximation used in deriving the expression for the flipping probability.

\section{Acknkowledgments}
We thank Guido Montagna for feedback on preliminary versions of this manuscript. F.D.D. gratefully acknowledges the Department of Computer Science at University College London for their hospitality, and the Erasmus Traineeship Program for financial support.

\bibliography{bibliography}
\end{document}